\documentclass[aps,prb,twocolumn,groupedaddress,showpacs,showkeys,amsmath]{revtex4-1}
\usepackage[dvips]{graphicx}
\usepackage{dcolumn}
\usepackage{bm}
\usepackage{color}
\usepackage{longtable}
\usepackage{ctable}
\usepackage{rotating}
\usepackage{booktabs}

\begin{document}

\title{Unconventional phase III of high-pressure solid hydrogen} 

\author{Sam Azadi}

\affiliation{Department of Physics, King’s College London, Strand, London WC2R 2L,
Department of Physics, Imperial College London, Exhibition Road, London SW7 2AZ,
United Kingdom}
\email{sam.azadi@kcl.ac.uk}

\author{Thomas D. K\"{u}hne}

\affiliation{Chair of Theoretical Chemistry and Paderborn Center for Parallel Computing,
University of Paderborn, Warburger Str.\ 100, D-33098 Paderborn, Germany}

\date{\today}

\begin{abstract}
We reassess the phase diagram of high-pressure solid hydrogen using mean-field and 
many-body wave function based approaches to determine the nature of phase III of solid hydrogen.
To discover the best candidates for phase III, density functional theory calculations within the meta-generalized gradient approximation by means of the 
strongly constrained and appropriately normed (SCAN) semilocal density functional are employed. 
We study eleven molecular structures with different symmetries, which are the most competitive phases, 
within the pressure range of 100 to 500~GPa.
The SCAN phase diagram predicts that the $C2/c-24$ and $P6_122-36$ structures
are the best candidates for phase III with  an energy difference of less than 1~meV/atom. 
To verify the stability of the competitive insulator structures of $C2/c-24$ and $P6_122-36$, 
we apply the diffusion Monte Carlo (DMC) method to optimise the
percentage $\alpha$ of exact-exchange in the trial many-body wave function. 
We found that the optimised $\alpha$ equals to $40 \%$, and denote the corresponding exchange and correlation functional as PBE1. 
The energy gain with respect to the well-known hybrid functional PBE0, where $\alpha = 25\%$, varies
with density and structure. The PBE1-DMC enthalpy-pressure phase diagram
predicts that the $P6_122-36$ structure is stable up  to 210~GPa, where 
it transforms to the $C2/c-24$. Hence, we predict that the phase III of high-pressure 
solid hydrogen is polymorphic. 
\end{abstract}

\maketitle

\section{Introduction}

The  phase diagram of high-pressure hydrogen is a challenging problem in condensed matter and high-pressure physics.
It has been extensively studied since 1935 \cite{1935} by experiment, theory and more recently computational methods.
The main interests are  the relevance of solid metallic hydrogen to room-temperature 
superconductivity\cite{Ashcroft}, possible existence of a metallic liquid ground state\cite{Bonev,JChen},
 and astrophysics\cite{Hemley,Ginzburg, RevMod12}. 

In this work, we focus on low temperature phases
of solid hydrogen. Infrared (IR) and Raman measurements 
suggest the existence of several phases and phase transitions within the low 
temperature region of the phase diagram.
Phase I, which is stable up to 110$\pm$5~GPa, is
a molecular solid composed of quantum rotors arranged in a hexagonal
close-packed structure. Changes in the low-frequency regions of the
Raman and IR spectra imply the existence of phase II, also known
as the broken-symmetry phase, above 110$\pm$5~GPa.  The appearance of
phase III at $\sim$150~GPa is verified by a large discontinuity in the
Raman spectrum and a strong rise in the spectral weight of the molecular
vibrons\cite{Hemley}.
The IR activity increases dramatically upon transition
from phase II to phase III, and  the  IR  and  Raman  vibron  frequencies 
soften  by  about 80 $(1/cm)$ \cite{XZLi}. 
Phase IV, characterized by two vibrons in its Raman
spectrum, was discovered at 300 K and pressures above 230
GPa\cite{Eremets, Howie, Howie2}. A new phase has been observed at
pressures above 200~GPa and higher temperatures (for example, 480 K at
255~GPa)\cite{Howie3}. This phase is thought to meet phases I and IV
at a triple point, near which hydrogen retains its molecular
character. The most recent experimental results\cite{Simpson} indicate
that H$_2$ and hydrogen deuteride at 300 K and pressures greater than
325~GPa transform to a new phase V, characterized by substantial
weakening of the vibrational Raman activity. The structure of 
high-pressure solid hydrogen above 150~GPa is experimentally unknown. 
Thus, theoretical and computational techniques play a crucial role 
in determining the structure of solid hydrogen. The main goal of 
this work is to determine the phase III of solid hydrogen using the 
most accurate {\it ab-initio} techniques. 

The electronic structure properties and lattice dynamic of solid
hydrogen were mainly investigated using density functional theory
(DFT) with local and semi-local exchange-correlation (XC) functionals
\cite{Pickard,Pickard2,Goncharov,Magdau,Naumov,Morales2013,Clay,Clay16,JETP,singh,PRB13}
including van der Waals XC functionals \cite{PCCP17}.
In particular, DFT with generalized gradient approximation (GGA)
functionals were widely applied to search for low-energy
crystal candidate structures and to calculate their vibrational properties.
More accurate results for the phase diagram \cite{NJP,samprl,Neil15}, 
excitonic and quasi-particle band gaps of insulator phases\cite{PRB17, JCC18}
were obtained by many-body wave function based 
quantum Monte Carlo (QMC) methods \cite{Matthew1,samjcp15,sambenz}.
These QMC results were used for benchmarking DFT functionals and it was 
indicated that the GGA for the XC functional can dramatically alter the
predicted phase diagram\cite{Clay16}.

The crystal structure is the fundamental input for the first principle
calculations.  Due to lack of established experimental results for the 
crystal structure, there is no option but to use the structures predicted by
DFT. Most of the structures have been predicted using the
Perdew-Burke-Ernzerhof (PBE)\cite{PBE} XC
functional\cite{Pickard,Pickard2}. Yet. it is now widely
accepted that DFT results for high-pressure hydrogen strongly depends 
on the choice of employed XC functional\cite{Clay,PRB13}.
To the best of our knowledge, there is no comparison between the predicted 
structures by DFT-based structure prediction methods, 
which are performed using different XC functionals. 
The QMC results rely on the structures that are obtained 
by DFT simulations since structure prediction by QMC is yet unaffordable 
due to computational costs of dealing with many-body wave functions. 
Therefore, benchmarking DFT functionals and 
calculating the phase diagram using different XC functionals are 
important for finding the most accurate XC functional. Moreover,
the functionality of the enhancement factor within the high density 
regime and treatment of the exchange energy in DFT functionals 
yield different phase stability and phase transitions. 
Fortunately, there are some properties which are not affected by the XC 
approximation. For instance, recently we have proposed a rule of 
thumb of {\it the shorter the molecular bond-length, the larger the electronic band
gap and the higher the vibron frequencies}, which appears to be valid for all 
considered XC functionals \cite{PCCP17}. 

It was demonstrated that for the enthalpy-pressure (H-P) phase diagram calculations 
the best performing XC functionals over all densities are
the meta-GGA functionals\cite{Clay16}. However, the recently developed 
strongly constrained and appropriately normed (SCAN) meta-GGA 
functional\cite{scan15} has not been applied on solid hydrogen.
The behaviour of the exchange enhancement factor in the SCAN functional
as a function of the density gradient can provide more accurate 
results than using GGA functionals. Benchmarking DFT functionals 
indicated that the exchange energy and the exchange enhancement factor 
play a crucial role in H-P calculations \cite{Clay16}. In this work, 
we calculate the H-P phase diagram of molecular structures using
the SCAN XC functional.

The derivative discontinuity $\Delta_{XC}$ of the DFT-XC functional 
differs from the single-particle Kohn-Sham energy band gap. 
Local and semi-local DFT XC functionals underestimate the fundamental gap
because $\Delta_{XC}=0$ for them. The inclusion of Hartree-Fock exchange enables us to approximate 
the exchange contribution of $\Delta_{XC}$, but owing to the nonlocality, the exchange 
energy overestimates the band gap. Therefore, hybrid DFT, which includes a fraction of 
Hartree-Fock exchange, usually yields an improved description of the electronic structure of 
insulators. Moreover, because of the absence of an artificial self-repulsion between the 
occupied states, Hartree-Fock exchange cancels the self-interaction contribution of the Hartree energy 
and consequently provides a more accurate method to calculate the Kohn-Sham spectra of insulators than GGA. 
The fraction of Hartree-Fock exchange used in hybrid DFT can be considered as a variational parameters. 
In this work we optimize this variational parameter using the diffusion Monte Carlo (DMC) method 
to build an efficient variational many-body wave function, which accurately describes the electronic 
structural properties of insulators phases.

\section{Computational Details}\label{com}

In this work, we consider the molecular structures of $C2/c-24$, $P6_3/m-16$, 
$Pc-48$, $Pbcn-48$, $Cmca-24$, $Cmca-8$, $P6_122-36$, $Pca2_1-48$, $Pna2_1-48$, 
$C2-48$, and $P2_1/c-8$ within the pressure range of 100 to 500~GPa. The number 
after hyphen indicates the number of hydrogen atoms in the primitive cell used
in our calculations. Our DFT calculations were carried
out using the latest version of the Quantum-Espresso suite of programs\cite{QS}. 
We used a basis set of plane waves with an energy cutoff of 100~Ry. 
For the geometry and cell optimisations a 
$16\times16\times16$ ${\bf k}$-point mesh is employed for all the structures 
except of $Cmca-8$ and $P2_1/c-8$ for which a $24\times24\times24$ ${\bf k}$-point mesh
is used. The quasi-Newton algorithm was used for all cell and geometry 
optimization, with a convergence thresholds on the total energy and forces of 0.01 mRy and 0.1 mRy/Bohr,
respectively. 
To calculate the SCAN enthalpy-pressure phase diagram, 
we used a norm-conserving Troullier-Martin  pseudopotential without nonlinear  core  corrections, 
which was generated by the SCAN XC functional\cite{JCP-SCAN-PP}. It was demonstrated that 
use of other conventional pseudopotentials, which are not explicitly produced for the SCAN functional, 
can lead to discrepancies for some systems \cite{JCP-SCAN-PP}.
The relative enthalpy-pressure
phase diagram is simulated by fitting a quartic polynomial with five fitting parameters on nine 
enthalpy-pressure DFT points within the pressure range of 100 to 500~GPa. 

All QMC calculations were performed via  the CASINO package using a 
Slater-Jastrow trial wave function\cite{casino}. The orbitals of the Slater-determinant were obtained by 
means of DFT within the local density approximation using the developer version of the Quantum-Espresso code. 
The Slater orbitals were generated using a norm-conserving pseudopotential in conjunction with a basis set energy cutoff 
of 400~Ry that were transformed into a localized blip polynomial basis\cite{blip}. 
The PBE1 exchange-correlation functional,
in which the mixing of exact-exchange parameter $\alpha = 10, 25, 40, 60,$ and $80 \%$, was used to 
optimize the atomic coordinates and generate the single particle Kohn-Sham orbitals to be used 
in the QMC calculations. For the geometry optimization, the PBE pseudopotential is used. 
The DMC results were obtained using a real $\Gamma$-point wave function, $2\times2\times2$ super cell
size, and time step of 0.005 a.u. We used the the model-periodic Coulomb interactions \cite{MPC} to correct for Coulomb finite-size errors. 
We used the conventional Jastrow factor that includes the polynomial
one-body electron-nucleus, two-body electron-electron, and three-body electron-electron-nucleus terms, 
which were optimized by minimizing the variance at the VMC level\cite{variance}. 

\section{Results and discussion}
\subsection{SCAN phase diagram}

\begin{figure}
\includegraphics[width=0.5\textwidth]{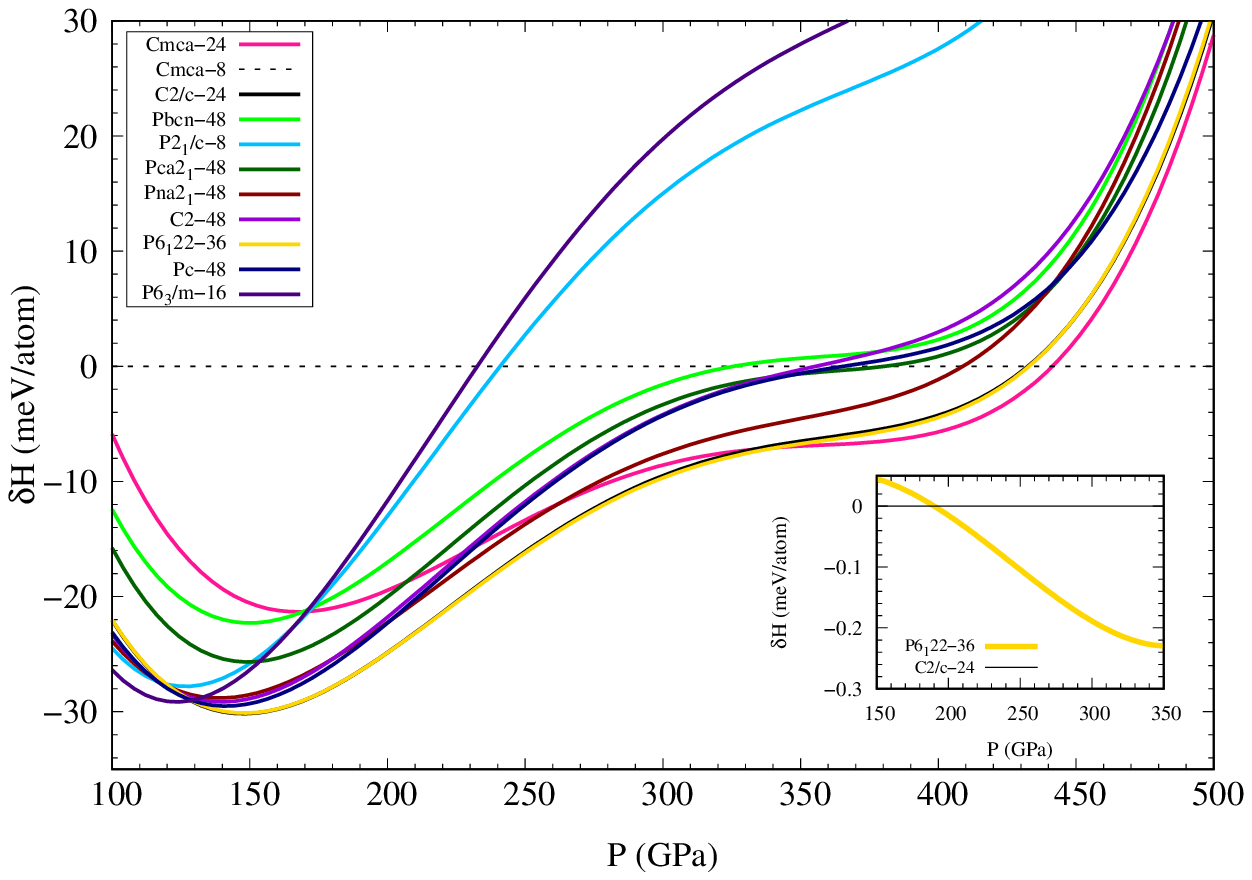}
\caption{\label{HP} (Color online) The static relative enthalpy-pressure phase diagram for solid 
molecular hydrogen as calculated by the SCAN XC functional. There are four phase transitions: 
$P6_3/m-16$ to $C2/c-24$ at 129~GPa,  $C2/c-24$ to $P6_122-36$ at 190~GPa, 
$P6_122-36$ to $Cmca-24$ at 343~GPa, and $Cmca-24$ to $Cmca-8$ at 442~GPa. 
The number after the hyphen indicates the number of hydrogen atoms in the primitive cell used
in our DFT calculations.}  
\end{figure}
In the following, we first discuss the predicted DFT-SCAN phase diagram.
Fig.~\ref{HP} illustrates the H-P phase diagram for the molecular structures 
of solid hydrogen within the pressure range of 100 to 500~GPa, which is simulated using 
the SCAN functional. 
The studied structures were predicted in previous works by {\it ab-initio}
random structure searching method (AIRSS) using GGA functionals at the static 
level in which the vibrational contributions were not included \cite{Pickard, Pickard2}. 
The SCAN H-P phase diagram predicts four phase transitions of hexagonal $P6_3/m-16$ to 
monoclinic $C2/c-24$ at 129~GPa, $C2/c-24$ to hexagonal $P6_122-36$ at 190~GPa, 
$P6_122-36$ to orthorhombic $Cmca-24$ at 343~GPa, and $Cmca-24$ to $Cmca-8$ at 442~GPa.
The hexagonal $P6_3/m-16$ structure is the most stable phase at pressures below $\sim$130~GPa and is the 
best candidate for the phase I of solid hydrogen. This agrees with low pressure static phase diagram 
as calculated by GGA functionals \cite{Pickard, JETP, PRB13}. The other candidate for the pressure 
range below 150~GPa is $P2_1/c$ which was initially predicted with eight atoms per primitive cell. 
Previous static DFT calculations using different GGA functionals, where the $P6_3/m-16$
were not considered, suggested that the $P2_1/c-24$ with twenty four hydrogen atoms per
primitive cell is the best candidate for phase I \cite{Neil15}. 

According to the SCAN phase diagram, $C2/c-24$ is stable within the pressure window of 129-190~GPa
and it transforms to $P6_122-36$ at 190GPa. 
The $P6_122-36$ phase has been predicted by AIRSS with the Becke-Lee-Yang-Parr (BLYP)\cite{BLYP}
XC functional \cite{tomeu16}. According to the static DFT-BLYP phase diagram the $C2/c-24$ is more stable 
than the $P6_122-36$ phase within pressure range of 100-350~GPa, whereas the $P6_122-36$ phase stabilizes because of the 
zero point energy (ZPE) contributions. The DFT-BLYP Gibbs free energy phase diagram at zero temperature, where 
the ZPE is taken into account, suggested that $P6_122-36$ is more stable than the $C2/c-24$ phase at pressures
below 180~GPa \cite{tomeu16}. Our static SCAN phase diagram predicts that the $P6_122-36$ structure is the best candidate for 
phase III of solid hydrogen above $\sim$200~GPa and below $\sim$340~GPa. The $Cmca-24$ phase is stable from 
343~GPa till it transforms to $Cmca-8$ at 442~GPa. Our DMC calculations 
indicated a very similar phase transition of $Cmca-24$ to $Cmca-8$ at 439~GPa \cite{samprl}. All previous 
studies of the high-pressure solid hydrogen phase diagram, which were carried out using DFT based methods, 
predict that solid hydrogen adopts the $Cmca$ symmetry before atomization. Taking into account the consistent 
prediction of previous first-principles calculations within different levels of theory, we propose a new fact of 
{\it low-temperature solid hydrogen adopts a $Cmca$ symmetry before atomization}, which is independent of particular XC functional. 

\begin{figure}
\includegraphics[width=0.40\textwidth]{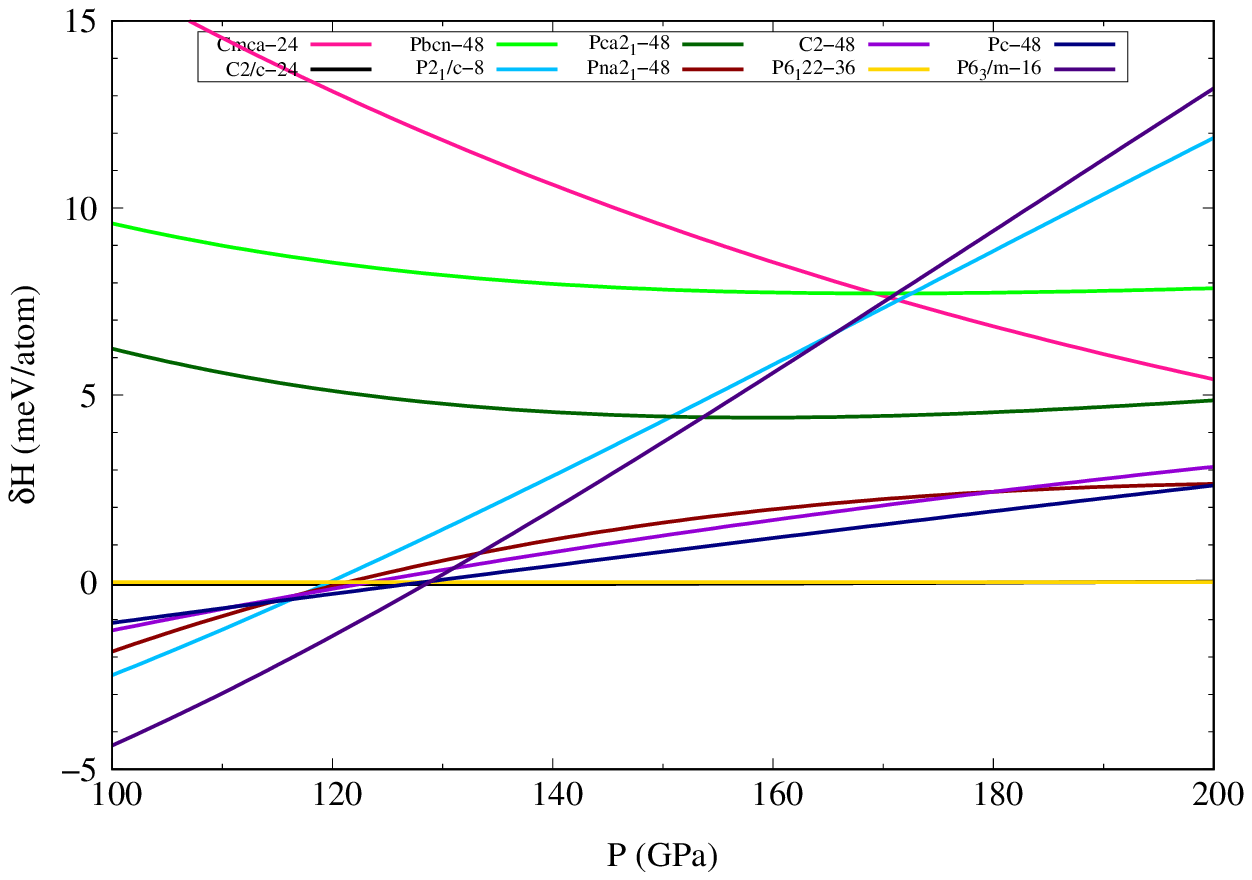}\\
\includegraphics[width=0.40\textwidth]{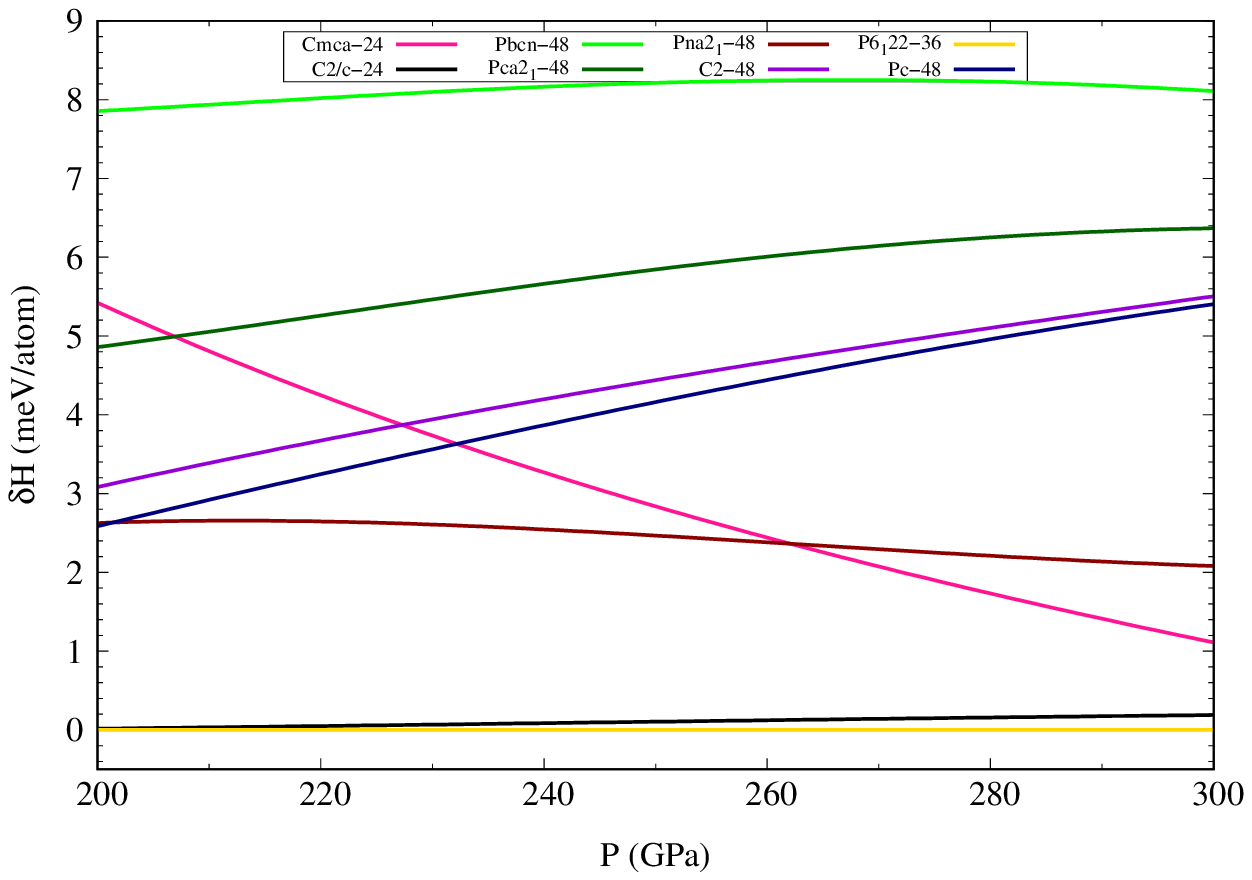}\\
\includegraphics[width=0.40\textwidth]{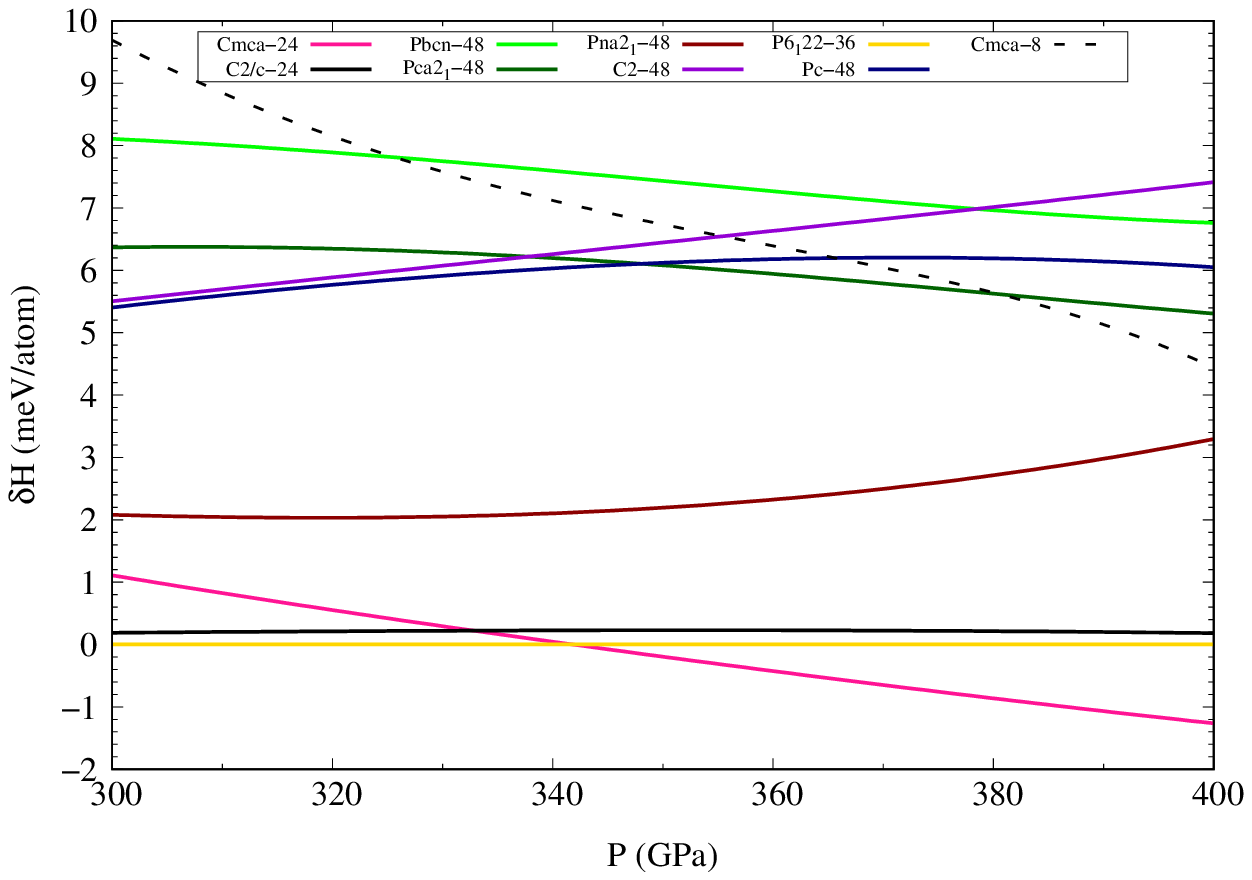}\\
\includegraphics[width=0.40\textwidth]{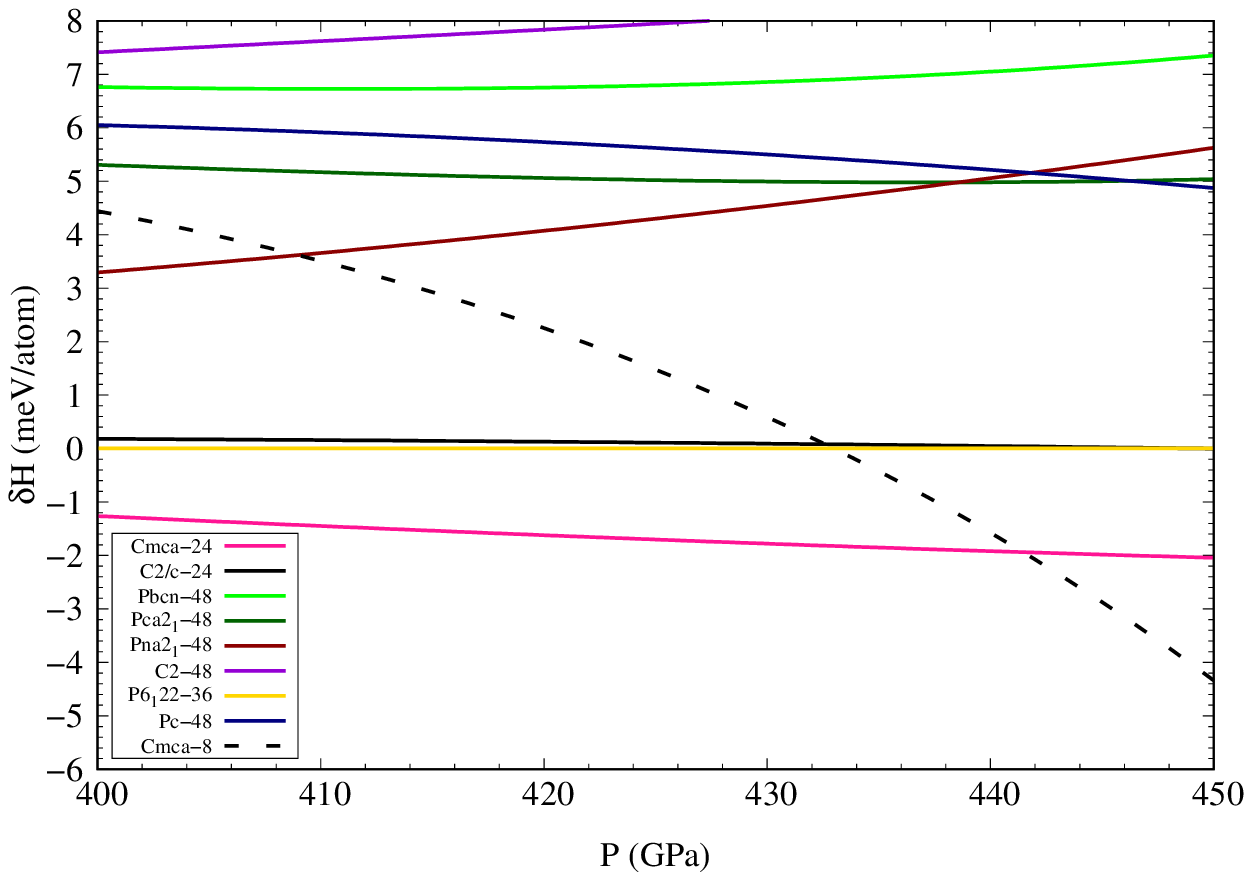}\\
\caption{\label{HP_sep} (Color online) The static relative enthalpy-pressure 
	phase diagram of solid molecular hydrogen. The phase diagram 
	is separated into four pressure ranges of 100-200, 200-300, 300-400 and 
	400-450~GPa, respectively. The reference zero line corresponds to the $P6_122-36$ phase. The energy difference between 
	the $P6_122-36$  and $C2/c-24$ structures is less than 1~meV/atom. The observed phase transitions are:
	 $P6_3/m-16$ to $C2/c-24$ at 129~GPa,  $C2/c-24$ to $P6_122-36$ at 190~GPa, 
     $P6_122-36$ to $Cmca-24$ at 343~GPa, and $Cmca-24$ to $Cmca-8$ at 442~GPa.} 
\end{figure}

We studied the recently predicted structures of $Pca2_1-48$ and $Pna2_1-48$, 
which have been suggested as the best candidates for phase V at pressures above 300~GPa \cite{Tomeu18}. 
We found that none of these structure are stable at any pressure range. 
According to the BLYP phase diagram, at a pressure above 300~GPa\cite{Tomeu18}, 
$C2/c$ is stable until it transforms into the $Cmca-12$ phase at pressures above $\sim$420~GPa. 
The BLYP Gibbs free energy calculations predict that at low and room temperatures the $C2/c$ structure 
transforms into $Cmca-12$ at around 350~GPa. The static SCAN phase diagram indicates a phase transition 
around 343~GPa, namely the $P6_122$ to $Cmca-24$ phase transition. The BLYP dynamic phase diagram at zero temperature, 
where the ZPE contributions are included, showed the $P6_3/m$ to $C2/c$ phase transition at
130~GPa, which is close to the prediction of the static SCAN phase diagram. The reason could be due to the fact 
that the BLYP-ZPE difference between the $P6_3/m$ and $C2/c$ phases within the pressure range of 110-150~GPa
is smaller than 3~meV/atom\cite{PRB13}. 

\subsection{Diffusion Monte Carlo phase diagram}

Within the studied pressure range, the SCAN enthalpy difference between the $C2/c$ and $P6_122$ phases 
is lees than 1~meV/atom, which is below the accuracy of DFT. 
The DFT energy difference between these two phases within the pressure window of 160-260~GPa is around 0.3~meV/atom, 
which is affected by ZPE and thermal contributions. 
The IR and Raman spectra of $C2/c$ and $P6_122$ were 
calculated using DFT\cite{tomeu16}. Their 
results indicate that, since the $C2/c$ and $P6_122$ structures are identical, 
the frequencies of the active modes are indistinguishable and
agree with available experimental data. 
The only important difference between the two signals is the stronger IR vibron
peak for the $P6_122$ phase, which agrees with the experimental report that
in phase III the IR activity is much larger than that in phase II.
The main conclusion is that the IR and Raman spectra of $C2/c$ and $P6_122$
are consistent with the corresponding spectra observed for
phase III, which is why the structure of phase III can not be 
determined purely based on its vibrational response.
Hence, we apply a higher level of theory to determine phase III of solid hydrogen. 

In the rest of this work, we only focus on two structures of $C2/c$ ($C2/c-24$) and $P6_122$ ($P6_122-36$), 
which are the most likely candidates for phase III within the pressure range below 300~GPa and above 150~GPa. 
To calculate the phase transition between these two candidates and determining the structure of phase III, 
we employed the DMC method. Phase III of solid hydrogen has a finite energy band gap below 300~GPa, and the 
phase transition between the $C2/c$ and $P6_122$ structures is a pressure driven insulator-insulator structural
transition. For the DMC simulations, our initial aim is to find the trial many-body wave function that  
produces the best description of both phases. 
We generated a set of trail wave functions using the present DFT-PBE1 XC functional, 
in which the value of the exact-exchange parameter $\alpha$ is varied within the range of 0 to $80\%$.
The DMC approach is a variational method and therefore the $\alpha$ that gives the lowest ground state 
DMC energy provides the most accurate representation of the many-body  wave function of the system. 
We therefore use $\alpha$ to  vary the single particle orbitals of the trial wave function. Yet, prior to our DMC calculations, 
the atomic coordinates of each structure were fully relaxed for each value of $\alpha$. 

\begin{figure}[h!]
\begin{tabular}{c c}
\includegraphics[width=0.25\textwidth]{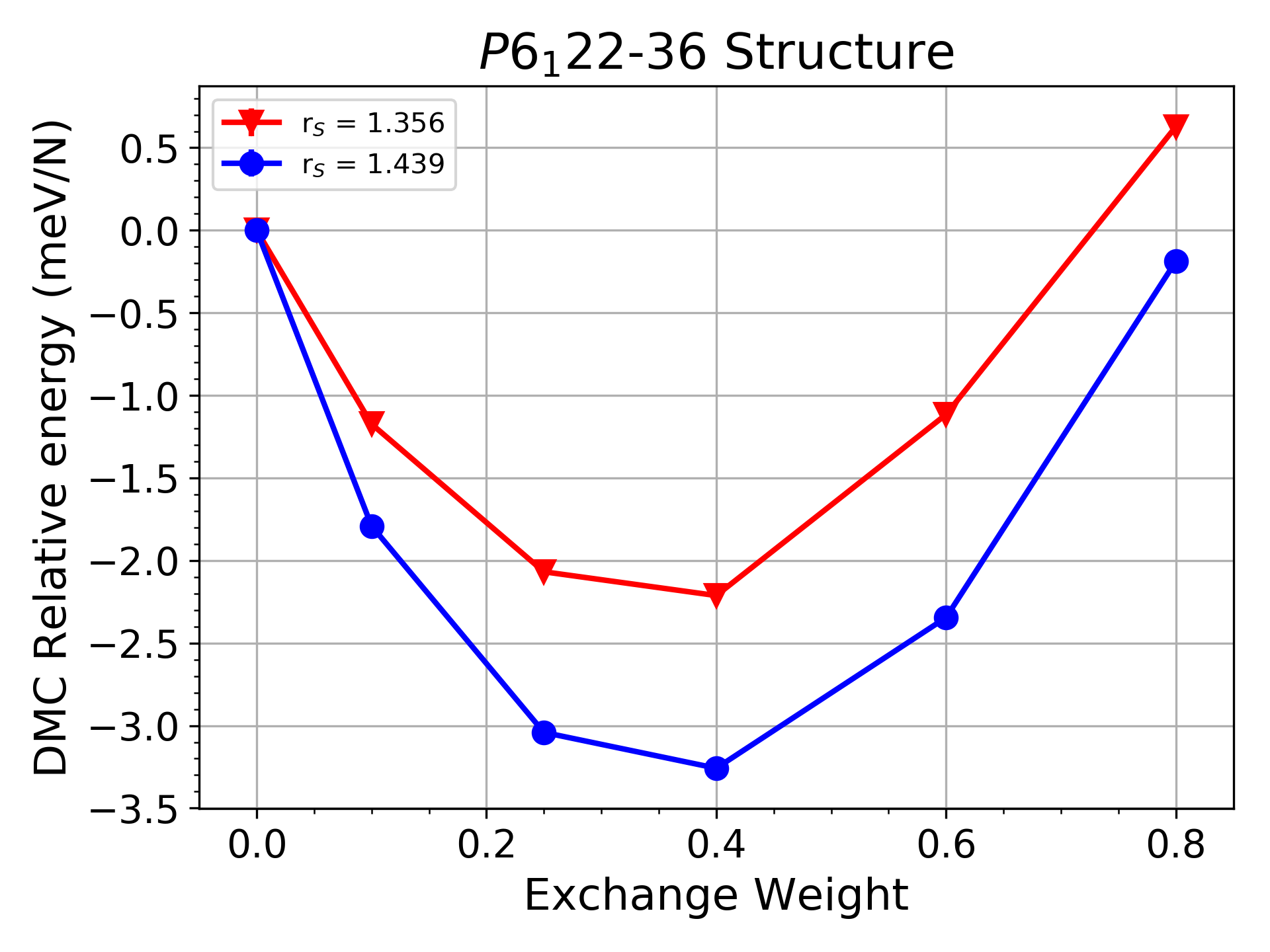} &
\includegraphics[width=0.25\textwidth]{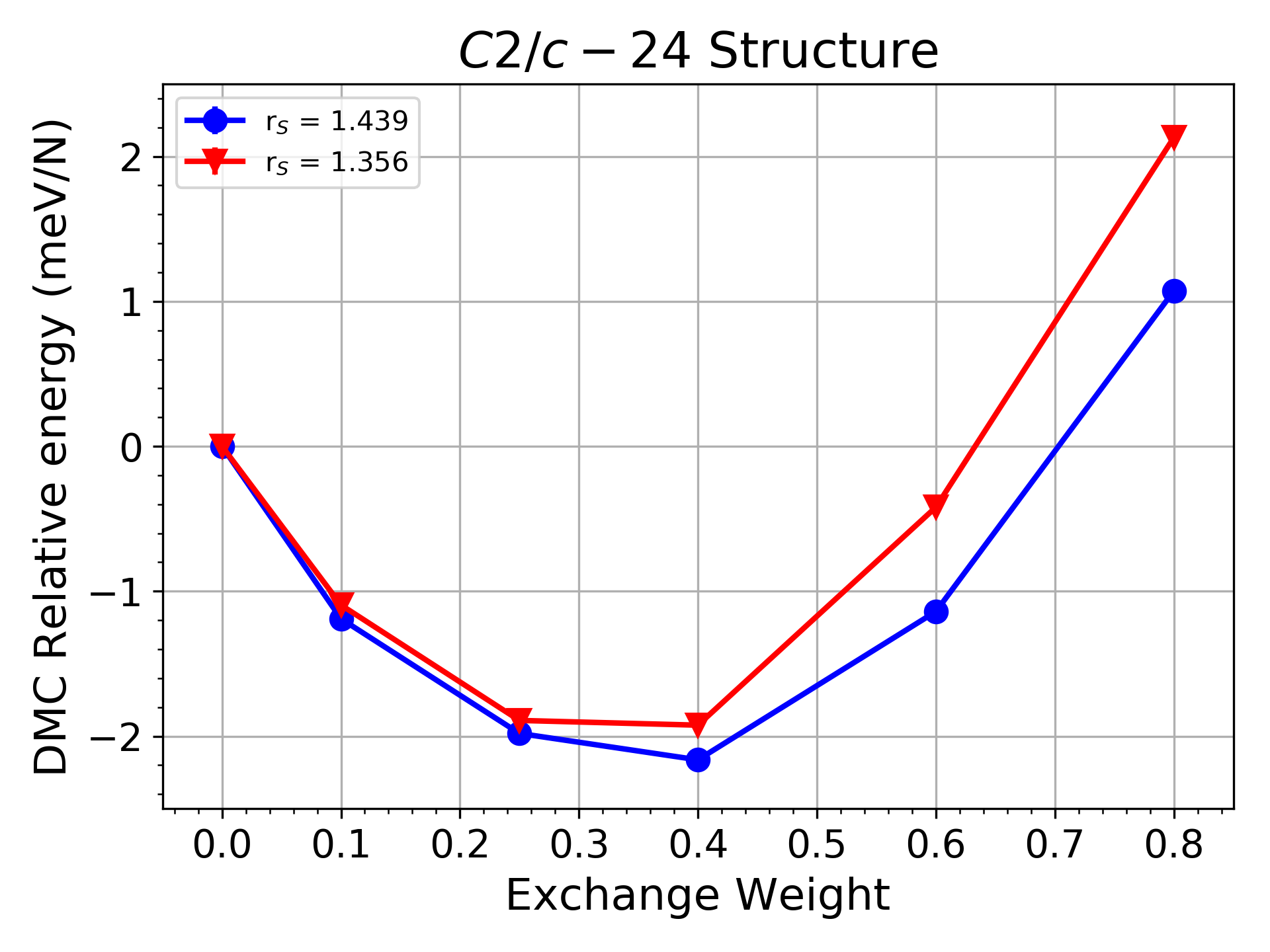}\\
\end{tabular}
\caption{\label{DMCX} DMC relative energy as a function of the percentage of exact-exchange $\alpha$ in the trial 
wave function. For $\alpha=0\%$, our PBE1 XC functional is identical to the PBE functional.
 The error bars are smaller than 0.1~meV/atom.}
\end{figure}

Fig.~\ref{DMCX} shows the DMC energies for the $C2/c$ and $P6_122$  structures as a function of
$\alpha$. The reference DMC energy was obtained using the 
conventional PBE exchange-correlation functional, i.e.$\alpha=0\%$.  The $C2/c$ and $P6_122$ phases 
were considered at two different densities with Wigner–Seitz radius of $r_S = 1.439$ and $1.356$ {\it a.u}, respectively. 
For the $P6_122$  structure a minimum in the DMC energy is observed for $\alpha = 40 \%$. 
The DMC energy difference between the  conventional PBE0 \cite{Perdew96, Becke93}, where $\alpha = 25 \%$, 
and our PBE1 with $\alpha = 40 \%$ at $r_S = 1.439$ and $1.356$ is 0.23(6), and 0.24(6) meV/atom.
In the case of the $C2/c$ structure, this difference 
at $r_S = 1.439$ and $1.356$ equals 0.18(5) and 0.06(5) meV/atom, respectively. The DMC results 
predict that $\alpha = 40 \%$ may tend to produce the best description of the ground state  electronic 
structure of the $P6_122$ and $C2/c$ structures, which are the most likely candidates for phase III 
of solid hydrogen. 
The energy gain at the minimum for $\alpha = 40\%$ with respect to the PBE XC functional for 
the $P6_122$ phase at $r_S = 1.439$ and $1.356$ is -3.26(4) and -2.21(3) meV/atom. 
For the $C2/c$ structure, the energy gain at the minimum with 
respect to PBE at $r_S = 1.439$ and $1.356$ equals to -2.16(2) and -1.92(3) meV/atom, respectively. Our DMC results 
indicate that for the $P6_122$ and $C2/c$ structures, reducing the $r_S$, which corresponds to higher densities
and consequently larger pressure, reduces the energy gain with respect to the PBE XC functional.

\begin{figure}[h!]
\begin{tabular}{c c}
\includegraphics[width=0.24\textwidth]{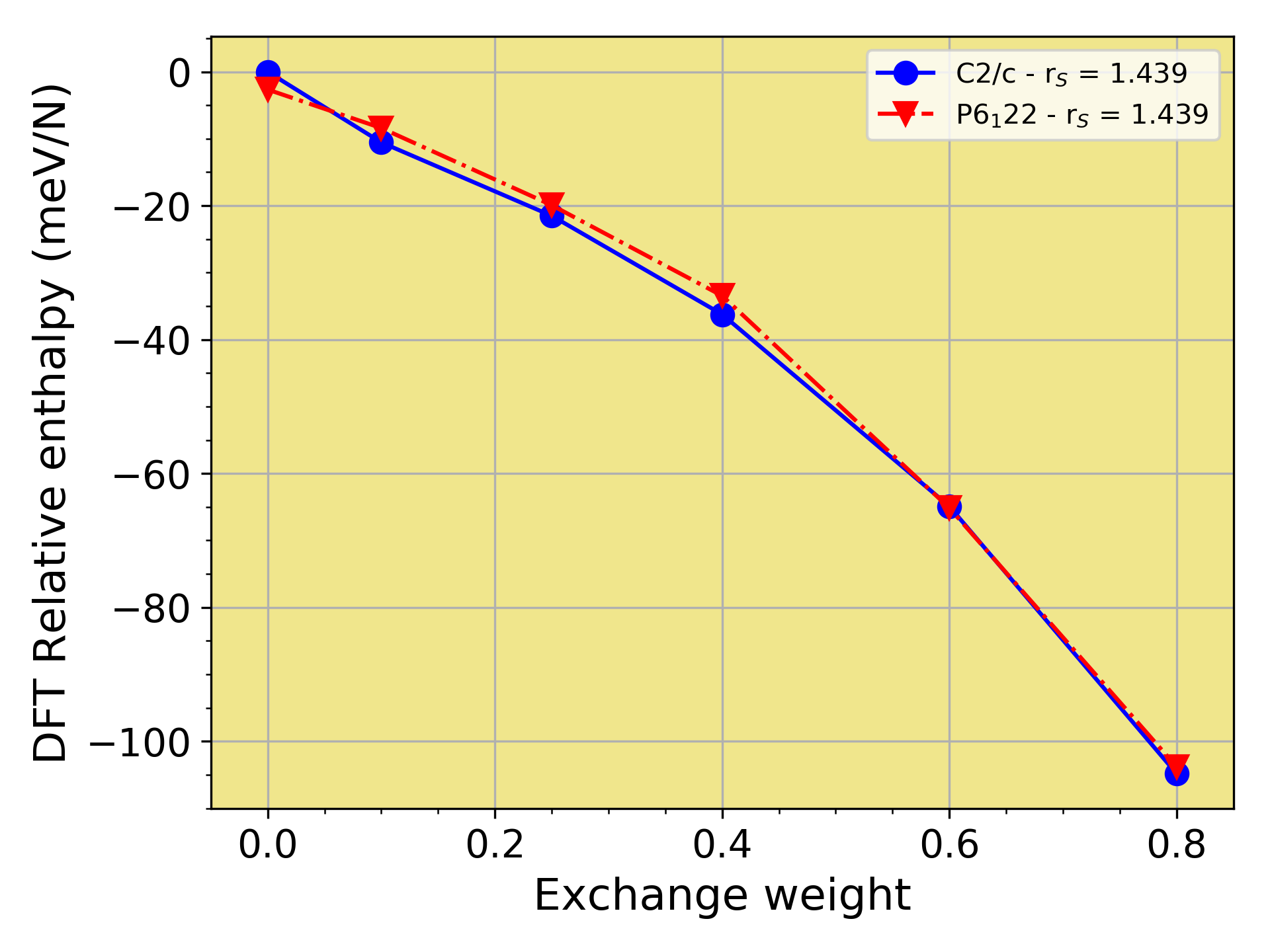} &
\includegraphics[width=0.24\textwidth]{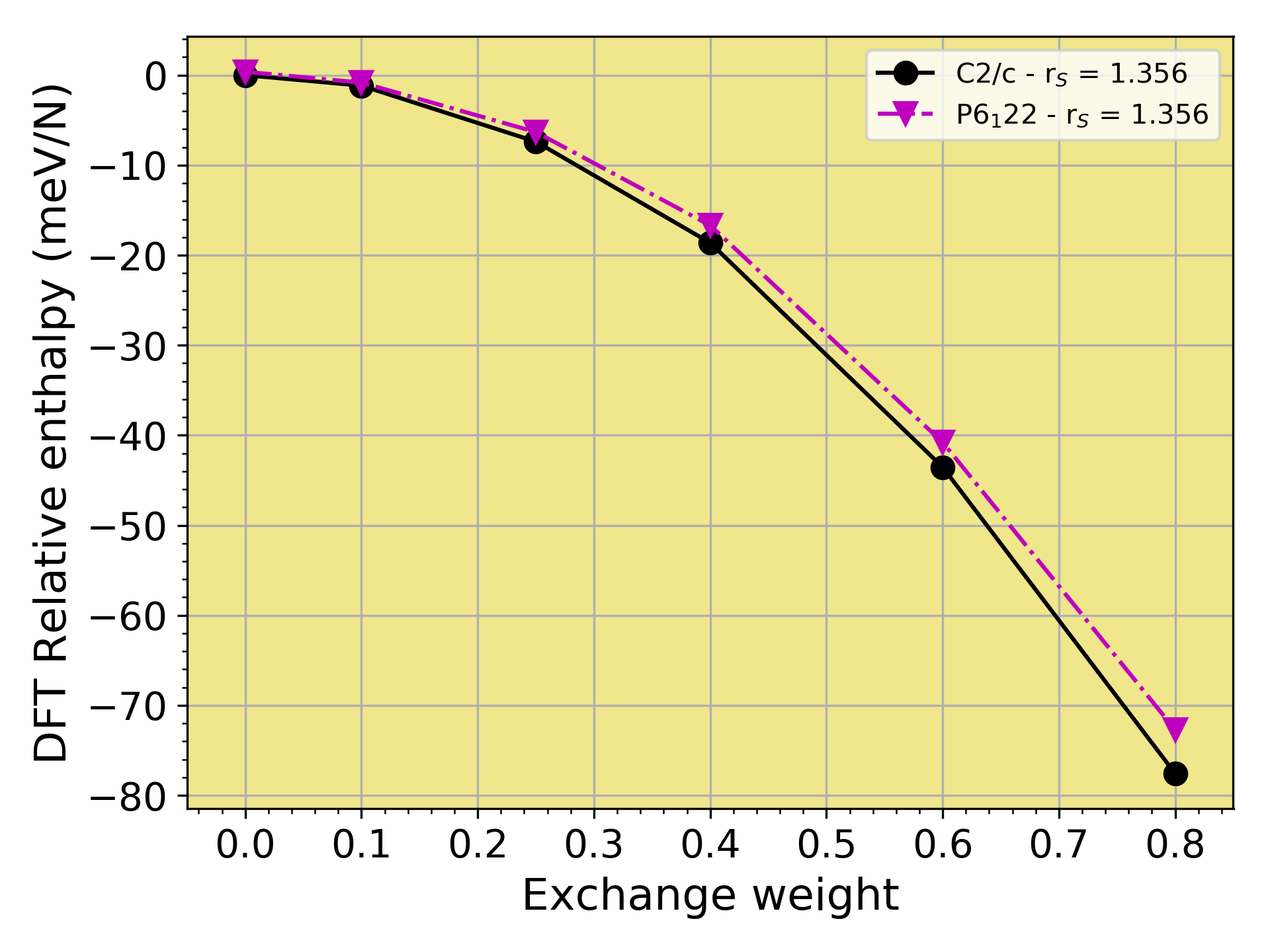} \\
\includegraphics[width=0.24\textwidth]{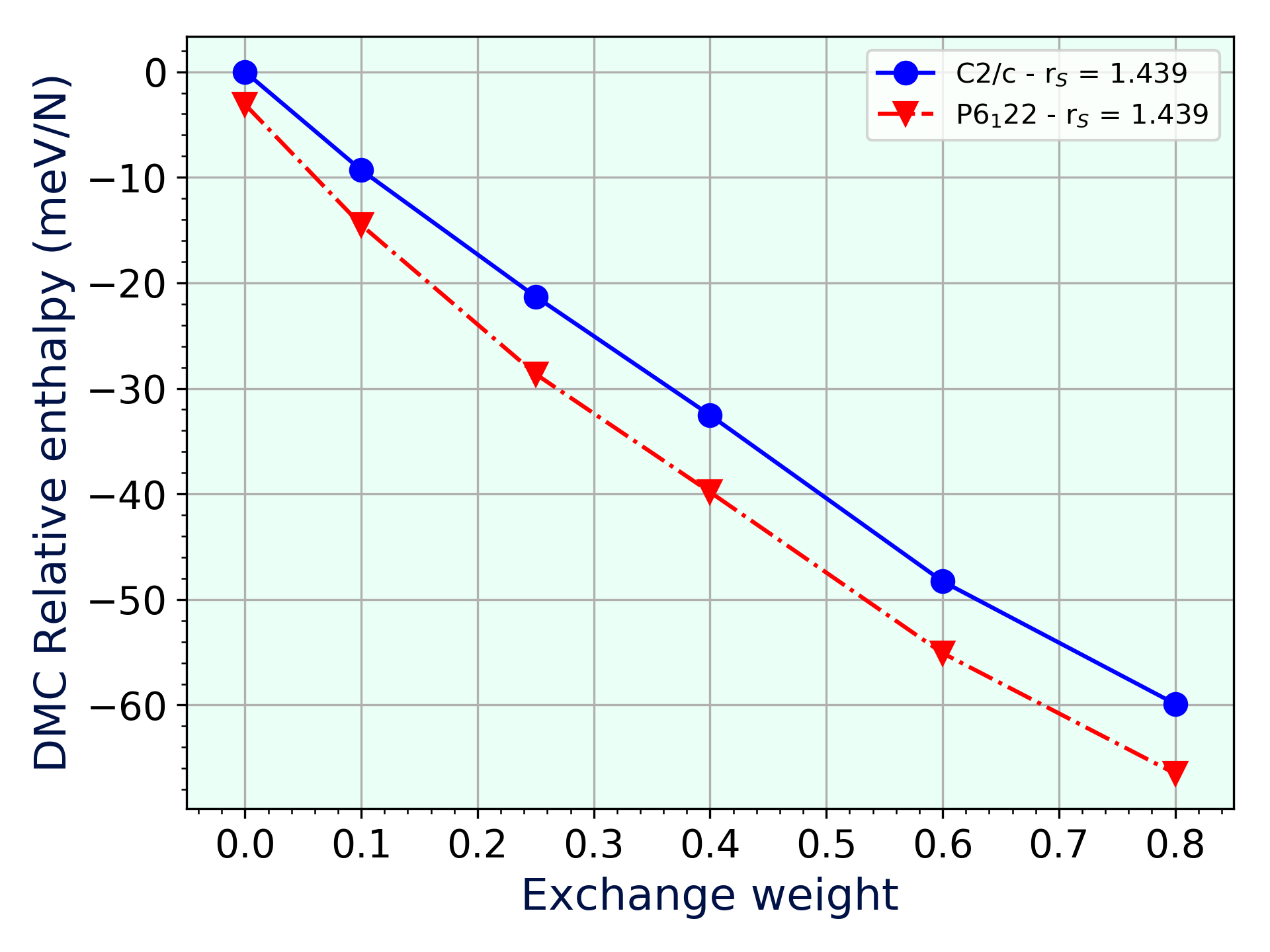}&
\includegraphics[width=0.25\textwidth]{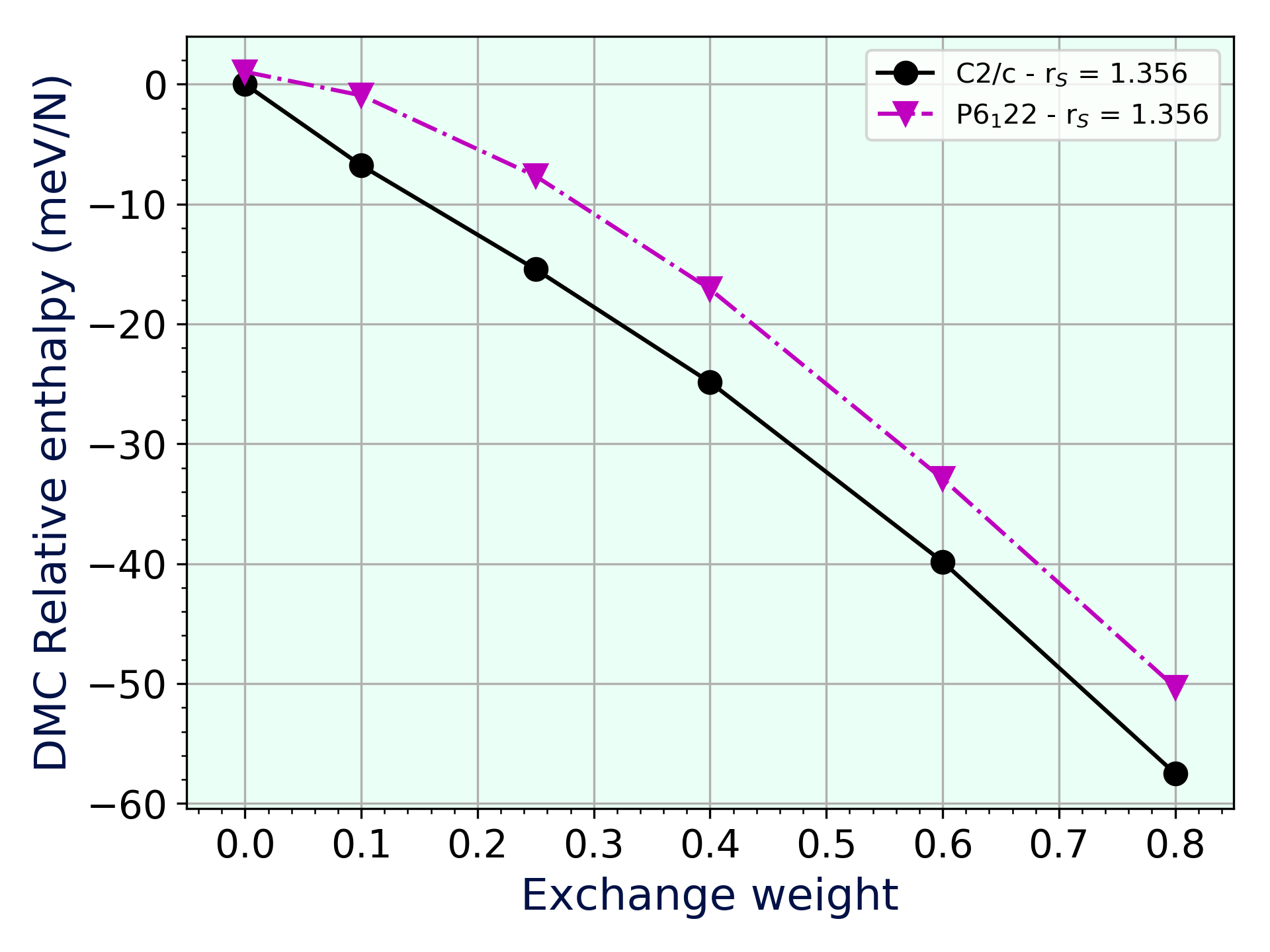}\\
\end{tabular}
\caption{\label{H_X} Relative DFT and DMC enthalpies of $C2/c$ and $P6_122$ structures as function of exchange weight. 
The reference point at each $r_S$ is the enthalpy of $C2/c$ phase.}
\end{figure}

In the Table~\ref{data_table} we have listed the pressure, the nearest neighbour distance (bond-length), and the energy band-gap 
for the $P6_122$ and $C2/c$ structures, which are obtained using the PBE ($\alpha=0\%$) and PBE1 ($\alpha=40\%$) XC functionals.
The difference between PBE and PBE1 pressures for the $C2/c$ and $P6_122$ phases at lower density is 3 and 2~GPa, respectively. 
The same pressure difference for the $C2/c$ and $P6_122$ structures at $r_S = 1.356$ is 5 and 4~GPa. Since the volume is 
fixed, the pressure difference between PBE and PBE1 functionals is only due to the atomic coordinate optimization. 
\begin{table}
\begin{tabular}{c c c c }
\hline\hline
 & $\alpha = 0$ ; $r_S = 1.439$ & \\ 
 \hline
Structure &  Pressure (GPa) & BL ($\AA$) & $E_g$ (eV) \\
$C2/c$ &      170 &  0.727 & 2.4    \\
$P6_122$ &  173 & 0.727 & 2.6 \\
\hline
 & $\alpha = 40 \%$ ; $r_S = 1.439$ & \\
 \hline
$C2/c$ &     167  &   0.717 & 5.7  \\
$P6_122$ & 171  &  0.716 & 5.9  \\
\hline 
& $\alpha = 0 ; r_S = 1.356 $ & \\
\hline
$C2/c$ & 262 & 0.730 & 0.92  \\
$p6_122$& 266 & 0.730 &  0.96  \\
\hline
& $\alpha = 40 \% ; r_S = 1.356 $ &\\
\hline 
$C2/c$ & 257 & 0.715 & 3.8 \\
$P6_122$ & 262 &  0.714 & 4.2   \\
\hline\hline
\end{tabular}
\caption{\label{data_table} The pressure in GPa, bond-length (BL) in $\AA$, and DFT energy band gap in (eV)  for the $C2/c$ and $P6_122$ structures, which are obtained at two densities of $r_S = 1.439$, and $r_S = 1.356$ using the PBE ($\alpha=0\%$) and PBE1 ($\alpha=40\%$) Xc functionals, respectively.}
\end{table}
We also used the PBE1 XC functional to calculate the enthalpy of $P6_122$ and $C2/c$ structures. 
The electronic structure  energy was obtained by DMC and the $PV$ term was calculated using the PBE1 functional. 
Fig.~\ref{H_X} illustrates the relative DFT and DMC enthalpies of the $C2/c$ and $P6_122$ structures
as a function of exchange weight $\alpha$ with respect to the PBE-$C2/c$ enthalpy. Similar to the SCAN prediction, the DFT-enthalpy phase diagrams, which are obtained for different values of $\alpha$, at $r_S = 1.439$ and $r_S = 1.356$ yield nearly 
indistinguishable results for the $C2/c$ and $P6_122$ structures. We found differences between the $C2/c$ and $P6_122$ phases 
by using the DMC electronic energy. Since $\alpha$ is not a variational parameter within hybrid-DFT, increasing 
the $\alpha$ reduces the Kohn-Sham total electronic energy for the systems with wide energy gaps. Our DMC simulations yields a 
parabola for the energy as a function of $\alpha$, where the energy gain of the $P6_122$ phase at lower densities is larger than that of the 
$C2/c$ one. For both studied phases, the DMC-enthalpy reduces with $\alpha$ due to contributions of the dominating DFT-$PV$
term, which is also decreasing with $\alpha$.

\begin{figure}[h!]
\includegraphics[width=0.40\textwidth]{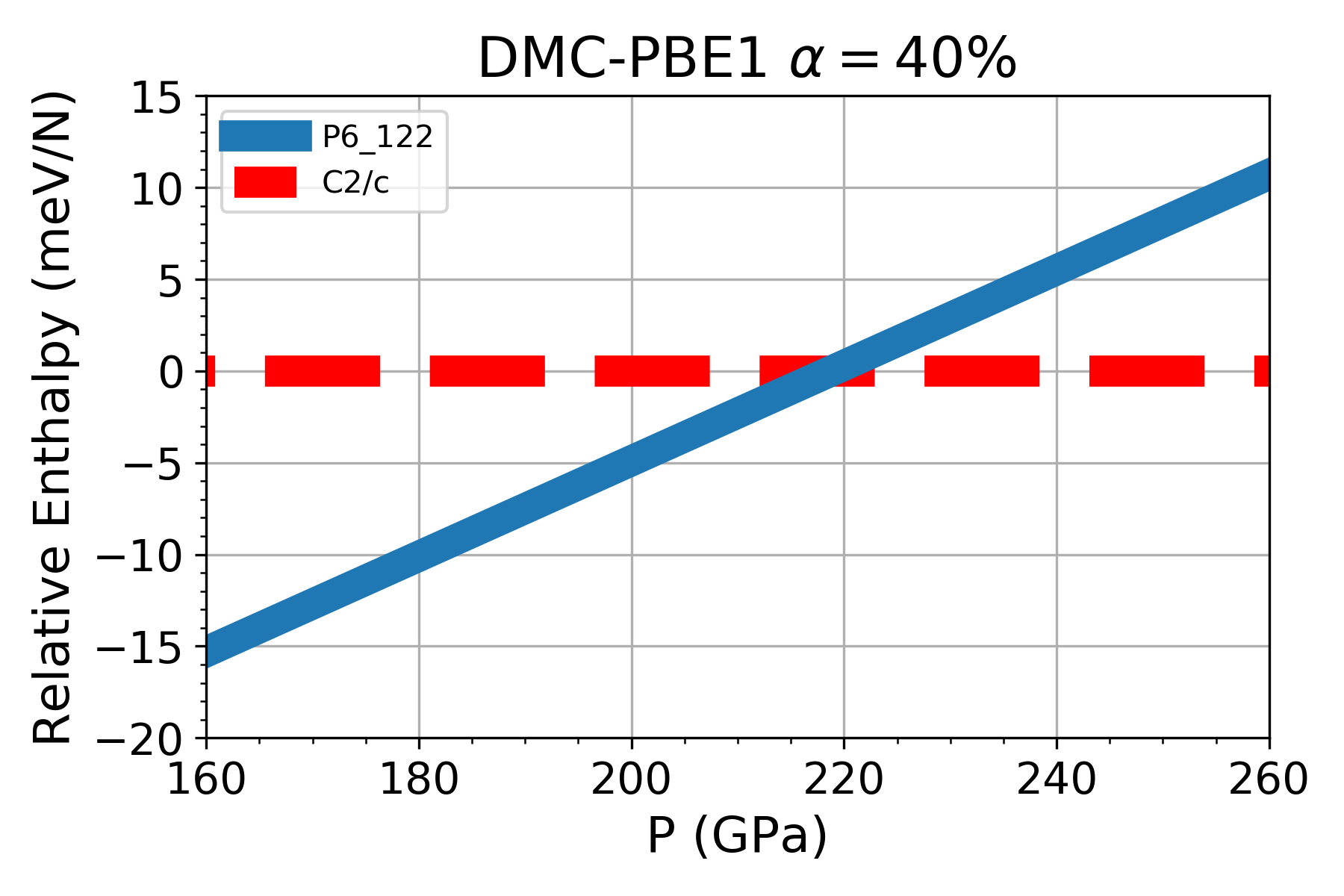} \\
\caption{\label{DMCHP} Relative stability of $P6_122$ with respect to the $C2/c$ structure. The enthalpy is obtained using the 
DMC energy and the DFT-PV term at $\alpha = 40 \%$. The $P6_122$ phase transforms into $C2/c$ at a pressure of 
around 210-220~GPa. The widths of the H-P lines indicates the estimated uncertainties in the enthalpy calculations, 
which is about $\sim$3 meV/atom. }
\end{figure}

Fig.~\ref{DMCHP} illustrates the relative enthalpy of $P6_122$ with respect to the $C2/c$ structure.
Our DMC H-P diagrams indicate that $P6_122$ is the stable phase within 
pressure range of 160-210~GPa and that it transforms to the $C2/c$ structure at around 210-220~GPa.
The $C2/c$ structure is the best candidate for the phase III at pressures above 210~GPa, whereas 
the $P6_122$ is the most likely candidate for phase III below 210~GPa. 
Our DMC phase diagram prediction is based on the DFT pressure, which introduces a 
$\sim$3-4 meV/atom error within the enthalpy\cite{NJP}. Obviously, using the DMC pressure 
for the static enthalpy-pressure phase diagram may alter the predicted outcomes.
The DMC phase diagram indicates that phase III of solid hydrogen could be polymorphic. 
Our prediction on the polymorphic nature of the phase III
agrees with recent nuclear magnetic resonance spectroscopy observations \cite{Tomeu19}, 
where the quantitative results are strongly depend on the used XC functional. 
A very recent experimental observation shows a first order phase transition near 425~GPa
from an insulating molecular phase to metallic hydrogen \cite{Loubeyre19}. The 
synchrotron infrared spectroscopy measurement indicates the stability of the insulating 
$C2/c-24$ phase in the high-density regime before matallization. Our DMC results for phase III 
and the stability of $P6_122$ at low-density and $C2/c$ at higher pressures agrees with 
this recent experiment.

Including the DFT-based ZPE contribution in our phase diagram calculations changes the phase 
stability of the phases and the transitions between them. We expect that including the phonon energies
will also change the SCAN H-P phase diagram. Note that the lattice dynamic should also be calculated using 
the SCAN functional, for instance by means of linear response theory or using the direct atom displacement method. 
The Gibbs free energy calculations up to room temperature for the $P6_122$ and $C2/c$ structures, 
which were performed by DFT-PBE and DFT-BLYP \citep{tomeu16}, indicate that the relative  phonon 
free energy within the pressure range of 150-300~GPa, in which the $P6_122$ to $C2/c$ phase transition occurs, 
is less than 2~meV/atom. Based on our DMC enthalpy pressure phase diagram, the energy difference 
between the $P6_1/22$ and $C2/c$ phases at 160 and 200~GPa are -15(3) and -5(3)~meV/atom, respectively. Therefore, we expect that 
the effects of including the ZPE and thermal contributions on the DMC H-P phase diagram are negligible.

\section{Conclusion}\label{con}

In the first part of this work, for the first time, we employed the
SCAN meta-GGA XC functional to provide a new revised static enthalpy-pressure 
phase diagram for eleven competitive molecular structures of solid hydrogen within the pressure range 
of 100-500~GPa. Our SCAN enthalpy-pressure phase diagram predicts four phase transitions of hexagonal $P6_3/m-16$ to 
monoclinic $C2/c-24$ at 129~GPa, $C2/c-24$ to hexagonal $P6_122-36$ at 190~GPa, 
$P6_122-36$ to orthorhombic $Cmca-24$ at 343~GPa, and $Cmca-24$ to $Cmca-8$ at 442~GPa.
Moreover, we compared all the available phase diagrams for high-pressure solid hydrogen, which were obtained 
by different DFT XC approximations and quantum Monte Carlo based methods, and proposed a 
new rule of thumb that {\it the molecular solid high pressure hydrogen obeys $Cmca$ symmetry before 
dissociating into an atomic physe.} We previously proposed our first rule of thumb about insulating 
molecular structures of high-pressure solid hydrogen\cite{PCCP17}, 
which is {\it the shorter the molecular bond-length the larger the electronic band gap and the higher vibron frequencies}.

In the second part of this work, we focused on determining the insulating phase III using the many-body 
wave function based DMC method. We considered two competitive candidates, namely the $C2/c$ and $P6_122$
structures that have finite energy band gaps. To find the most accurate trial many-body wave function, we 
took into account the fraction of exact-exchange $\alpha$ as a variational parameter, which was then 
optimised by DMC. We found that $\alpha = 40 \%$ is the optimised value that provides the lowest ground state 
electronic structure total energy. The energy gain with respect to the conventional PBE XC functional, where $\alpha=0\%$, 
depends on the atomic structure and density. Our DMC enthalpy-pressure phase diagram 
indicates that phase III of solid hydrogen adopts two structures, the one $P6_122$, which is 
stable below $\sim$210~GPa, and the $C2/c$ phase, which is stable at pressures above $\sim$210~GPa.

\section{Acknowledgements}
This project has received funding
from the European Research Council (ERC) under the European Union's Horizon
2020 research and innovation programme (grant agreement No 716142). 
The generous allocation of computing time by the Paderborn Center for Parallel 
Computing (PC$^2$) on OCuLUS and the FPGA-based supercomputer NOCTUA 
is kindly acknowledged.

\end{document}